# How to Answer Why - Evaluating the Explanations of AI Through Mental Model Analysis

Tim Schrills[†]
Institute for Multimedia and Interactive Systems
University of Lübeck
Lübeck, Germany
schrills@imis.uni-luebeck.de

Thomas Franke
Institute for Multimedia and Interactive Systems
University of Lübeck
Lübeck, Germany
franke@imis.uni-luebeck.de

## ABSTRACT

To achieve optimal human-system integration in the context of user-AI interaction it is important that users develop a valid representation of how AI works. In most of the everyday interaction with technical systems users construct mental models (i.e., an abstraction of the anticipated mechanisms a system uses to perform a given task). If no explicit explanations are provided by a system (e.g. by a self-explaining AI) or other sources (e.g. an instructor), the mental model is typically formed based on experiences, i.e. the observations of the user during the interaction. The congruence of this mental model and the actual systems functioning is vital, as it is used for assumptions, predictions and consequently for decisions regarding system use. A key question for human-centered AI research is therefore how to validly survey users' mental models.

The objective of the present research is to identify suitable elicitation methods for mental model analysis. We evaluated whether mental models are suitable as an empirical research method. Additionally, methods of cognitive tutoring are integrated. We propose an exemplary method to evaluate explainable AI approaches in a human-centered way.

## 1 INTRODUCTION AND BACKGROUND

Explainable AI (XAI) aims to improve various aspects of human-AI interaction, such as trust, traceability or predictability through explanations [1]. It can be assumed that all these factors are essentially influenced by the congruence of users' mental models and the actual system functioning (i.e., in the following labelled as 'physical model'). In the sense of cognitive science, a mental model can be understood as a set of knowledge elements that are interrelated [2]. From the existence of these elements and their connection, for example, the behavior of a system can be mentally simulated. These simulations are used to make predictions, i.e. assumptions about how the system will behave (in the case of an AI, for example, what classification will be made based on an input). This leads to the following conclusion: the higher the congruence between a mental model and the functioning of the system, the more accurate and correct the predictions [6]. Nevertheless, the number of correct predictions alone is not sufficient as a metric of an explanation's value, since it is not possible to deduce which errors in the mental model possibly cause prediction errors. In contrast, the extent of congruence (between users' models and physical model) before and after an explanation, interaction or other intervention can be used as a measure of an explanation's performance. For an adequate evaluation of systems - which also provides information on which knowledge elements are critical and can pose problems - a quantitative method for the elicitation of mental models is necessary. Hence, the development of a mental model analysis (MMA) is considered necessary [cf. 3].

Three different factors must be considered, which can lead to reduced congruence: (a) The absence of relevant knowledge elements; (b) the existence of irrelevant knowledge elements; and (c) the incorrect relation of knowledge elements. A simple example is a medical AI to classify presence of diabetes based on various factors. Potential users might not know, for example, that the AI also considers the time at which the diagnosis is requested, because for example meals have an effect on blood sugar level, (a) and thus gives a false prediction. Or users might assume that the AI integrates certain correlations, e.g. to assess blood glucose differently in diabetics than in healthy people - although the proportion of diabetics in the training data set may have been too low (c). While all these cases can be resolved post-hoc by an explanation, methods must be developed to reveal the differences between users' mental models and the physical model, e.g. to trigger specific explanations or to improve the presentation of the explanations.

In examining the impact of explanation, two objectives could be pursued: (1) It must be evaluated whether an explanation reduces the distance between the mental model of a user and the physical model. However, this is not always possible, since the physical model is (1a) not extractable or (1b) cannot be presented in a comprehensible way. (2) It is to be examined, which elements are influenced by an explanation. This could be important to understand whether an explanation is effective. The goal of MMA is thus not to determine whether a user has a better understanding of the physical model. Rather, the goal is to discover how knowledge elements are influenced by interventions, especially explanations. Combined with e.g. tasks concerning prediction or the subjective rating (by, using e.g. the System Causability Scale [4]), a better depiction of explanations' effects can be provided.

## 2 MENTAL MODEL ANALYSIS METHODS

We propose to structure existing methods into three different categories: The graphical representation of geothermal energy as in [9] is an example for (i) the mere collection of knowledge elements. Both, missing and irrelevant knowledge elements can be identified. By querying correlative relationships as in [5], an (ii) undirected relationship can be identified. Thus, besides (a) and (b), a subset of errors of (c) can be addressed. For a complete survey of mental models (i.e. a, b and c) it is necessary to identify (iii) directional relationships between knowledge elements. This can be done by collecting directed graphs (see [8]).

Additionally, we assume that the process of, for example, a classification, could be better surveyed by another format. A simple method for this could be the query of rules - but in what form can rules be queried? In the field of cognitive tutors, constraint-based tutoring systems [7] are used. These consist of a certain set of rules which are always composed of a relevance and a satisfaction clause. The first one triggers the rule, i.e. the circumstance that must be present for the fulfillment of the rule to be checked. In the example given above, this could be a heart disease. The latter represents a circumstance that must be present for the entire rule to be considered fulfilled, resulting in a simple if-then structure. For example, it could be that the blood pressure is ignored in the diagnosis. The advantage of this method is that the mental representation does not have to be modelled sequentially, but all existing, directed connections can be represented independently of each other. At the same time, the individual rules can be assumed to be simple enough to not cause a cognitive overload for participants.

## 3 PROPOSITION OF TEST PROCEDURE

Based on the analytical discussion above, we propose a procedure to apply constraint-based rules in the elicitation of mental models. For this purpose, participants are presented with different patient profiles and their symptoms. Furthermore, a classification of a Mock-Up AI is presented, which is based on clearly defined rules. This means that the rules that are to be applied are already known in advance to the researcher. This is different from the investigation of explanations in neural networks, for example, because they are unknown due to their nature as black boxes. To investigate the effect of an explanation it is not necessary that the physical model is given. However, by constructing such a system for research purposes it is possible to deliberately install problems that may occur in the real context (e.g. due to insufficient training data). After a defined number of observations, the participants are asked to define the rule set, according to which the AI makes its classification, by choosing relevance and satisfaction clauses. These consist of observable information from the patient profile. For example, the relevance clause could be that blood glucose is above a certain level, while the satisfaction clause is that a symptom is fatigue. Finally, it is assessed whether a fulfilled constraint increases or decreases the probability of classification. For example, in the given example, the classification "diabetes" might become more likely. A selection list is offered (for relevance, satisfaction and classification). The number of correct elements, missing elements and correct relations is then used to examine the congruence of the mental model and the physical model.

## 4 DISCUSSION AND CONCLUSION

We have proposed that the elicitation of mental models is a key step for the systematic improvement of XAI. Based on the proposed methodology, as a next step, the extent to which constraint-based rules are suitable to collect mental models and to quantitatively represent the changes triggered by e.g. explanations should be evaluated. This may open up a large and fruitful research space to AI researches, to create truly human centered AI that enables symbiotic cooperation and enhances the formation of explanation-based trust and acceptance.

## REFERENCES
[1] Ashraf Abdul, Jo Vermeulen, Danding Wang, Brian Y. Lim, and Mohan Kankanhalli. 2018. Trends and Trajectories for Explainable, Accountable and Intelligible Systems: An HCI Research Agenda. In *Proceedings of the 2018 CHI Conference on Human Factors in Computing Systems - CHI '18*, 1–18.
[2] Vladan Devedzic, Danijela Radovic and Ljubomir Jerinic. 1999. A Framework for Building Intelligent Manufacturing *Systems. IEEE Transactions on Systems Man and Cybernetics Part C (Applications and Reviews)*. 29. 422-439. 10.1109/5326.777077.
[3] Robert R. Hoffman, Shane T. Mueller, Gary Klein, and Jordan Litman. 2018. Metrics for explainable AI: Challenges and prospects. *arXiv preprint arXiv:1812.04608* (2018).
[4] Andreas Holzinger, André Carrington, and Heimo Müller. 2019. Measuring the Quality of Explanations: The System Causability Scale (SCS). Comparing Human and Machine Explanations. arXiv:1912.09024 [cs].
[5] John E Mathieu, Tonia S Heffner, and Gerald F Goodwin. The Influence of Shared Mental Models on Team Process and Performance. 11.
[6] D. A. Norman. 1987. Some Observations on Mental Models. In *Human-Computer Interaction: A Multidisciplinary Approach*. Morgan Kaufmann Publishers Inc., San Francisco, CA, USA, 241–244.
[7] Stellan Ohlsson. 1994. Constraint-based student modeling. In *Student modelling: the key to individualized knowledge-based instruction*. Springer, 167–189.
[8] Bingjun Xie, Jia Zhou, and Huilin Wang. 2017. How Influential Are Mental Models on Interaction Performance? Exploring the Gap between Users' and Designers' Mental Models through a New Quantitative Method. *Advances in Human-Computer Interaction* 2017, (2017), 1–14.
[9] Barbara S. Zaunbrecher, Johanna Kluge, and Martina Ziefle. 2018. Exploring Mental Models of Geothermal Energy among Laypeople in Germany as Hidden Drivers for Acceptance. *J. sustain. dev. energy water environ. syst.* 6, 3 (September 2018), 446–463.